# Absence of Long-Range Magnetic Ordering in a Trirutile High-Entropy Oxide $(Mn_{0.2}Fe_{0.2}Co_{0.2}Ni_{0.2}Cu_{0.2})Ta_{1.92}O_{6-\delta}$

*Gina Angelo,[a] Liana Klivansky,[b] Jeremy G. Philbrick,[c] Tai Kong,[c,d] Jian Zhang,[b] Xin Gui [a*]*

[a] Department of Chemistry, University of Pittsburgh, Pittsburgh, PA, 15260, USA
[b] The Molecular Foundry, Lawrence Berkeley National Laboratory, Berkeley, CA, 94720, USA
[c] Department of Physics, University of Arizona, Tucson, AZ, 85721, USA
[d] Department of Chemistry and Biochemistry, University of Arizona, Tucson, AZ, 85721, USA

***Abstract***

Functionalities of solid-state materials are usually considered to be dependent on their crystal structures. The limited structural types observed in the emerged high-entropy oxides put constraints on exploration of their physical properties and potential applications. Herein, we synthesized the first high-entropy oxide in a trirutile structure, $(Mn_{0.2}Fe_{0.2}Co_{0.2}Ni_{0.2}Cu_{0.2})Ta_{1.92}O_{6-\delta}$, and investigated its magnetism. The phase purity and high-entropy nature were confirmed by powder X-ray diffraction and energy-dispersive spectroscopy, respectively. X-ray photoelectron spectroscopy indicated divalent Mn, Co, Ni and Cu along with trivalent Fe. Magnetic properties measurements showed antiferromagnetic coupling and potential short-range magnetic ordering below ~ 4 K. The temperature-dependent heat capacity data measured under zero and high magnetic field confirmed the lack of long-range magnetic ordering and a possible low-temperature phonon excitation. The discovery of the first trirutile high-entropy oxide opens a new way for studying the relationship between the highly disordered atomic arrangement and their magnetic interaction. Furthermore, it provides a new direction for exploring functionalities of high-entropy oxides.

## Introduction

High-entropy oxides (HEOs) are a type of material where five or more elements are randomly distributed in equiatomic stoichiometry on the same atomic site. Vast applications such as catalysis[1] and reversible energy storage[2] were found for HEOs. Better mechanical properties and resistance to oxidation/corrosion were also observed for HEOs[3,4]. In addition, magnetic insulating HEOs were reported due to their potential application in the next-generation memory and spintronic devices[5]. Since Rost et. al. discovered the first HEO in rocksalt structure[6], there have been great efforts to expand the structural types of HEOs. To date, ten major crystal structures were found for HEOs: bixbyite ($A_2O_3$)[7], delafossite ($ABO_2$)[8], fluorite ($AO_2$)[9], magnetoplumbite ($AB_{12}O_{19}$)[10], mullite-type ($A_2B_4O_{10}$)[11], perovskite ($ABO_3$)[12], pyrochlore ($A_2B_2O_7$)[13,14], rocksalt (AO)[6], Ruddlesden-Popper phase ($A_{n+1}B_nO_{3n+1}$)[15] and spinel ($AB_2O_4$)[16] where A and B are cations. However, many structural types remain obscure, which significantly limits the manipulation and exploration of HEOs' properties. Furthermore, research surrounding their magnetic properties is only recently studied[17]. Upon investigation, HEOs can possess complicated magnetic properties due to the disordering of multiple transition metals, e.g., the spin-glass states in perovskite HEOs[18] and tunable magnetism in spinel HEOs[16]. Considering the importance of crystal structure in determining physical properties, expanding the structural families of HEOs is crucial for exploring their functionalities and potential applications.

Herein, we report the first trirutile HEO, $(Mn_{0.2}Fe_{0.2}Co_{0.2}Ni_{0.2}Cu_{0.2})Ta_{1.92}O_{6-\delta}$, belonging to space group $P\ 4_2/mnm$ (no. 136) evidenced by X-ray diffraction (XRD) and energy-dispersive spectroscopy (EDS). While there were reports of rutile HEOs[19], the trirutile structure remains unfounded. Trirutile i.e. $AB_2O_6$, is a derivative of the rutile structure i.e. $AO_2$, wherein an additional element disrupts the repeatability of the rutile structure such that a 1×1×3 superlattice is formed. $NiTa_2O_6$[20], $CuTa_2O_6$[21] and $CoTa_2O_6$[22] have all been studied for their antiferromagnetic properties as trirutile oxides and we employed $CoTa_2O_6$ as the parent compound, which was reported to be antiferromagnetically ordered below $T_N \sim 6.6$ K[22]. However, when doped with Mg, i.e., $Co_{1-x}Mg_xTa_2O_6$, the antiferromagnetic ground state is quenched at x=10% while a short-range ferromagnetic correlation can be found in all series of compositions[23]. Considering that the theoretical calculations suggest the ferromagnetic ground state[23] competing ferromagnetic and antiferromagnetic interactions must exist in $CoTa_2O_6$. Our investigation on the physical properties of $(Mn_{0.2}Fe_{0.2}Co_{0.2}Ni_{0.2}Cu_{0.2})Ta_{1.92}O_{6-\delta}$ suggests short-range antiferromagnetic interaction resulting from the highly disordered transition-metal site. Therefore, the trirutile $CoTa_2O_6$ provides a good platform for investigating the interplay between magnetism and the crystallographic disorder on the Co site, with further possibilities to extend the scope of physical properties in trirutile HEOs.

*Experimental Details*

**Synthesis of ($Mn_{0.2}Fe_{0.2}Co_{0.2}Ni_{0.2}Cu_{0.2}$)$Ta_{1.92}O_{6-\delta}$**

($Mn_{0.2}Fe_{0.2}Co_{0.2}Ni_{0.2}Cu_{0.2}$)$Ta_{1.92}O_{6-\delta}$ was prepared by mixing CoO powder (99.995%, Thermo Scientific), MnO powder (99.99%, Thermo Scientific), CuO powder (99.7%, ~200 mesh, Alfa Aesar), NiO powder (99.998%, Thermo Scientific), $Fe_2O_3$ powder (99.9%, Thermo Scientific), and $Ta_2O_5$ powder (99.5% Thermo Scientific), and by placing the 0.2 : 0.2 : 0.2 : 0.2 : 0.1 : 0.96 mixture into an alumina crucible. The crucible was covered with a small alumina lid, placed into a larger crucible, and then surrounded with activated charcoal (untreated, ≤5 mm, Sigma-Aldrich). This assembly was then placed into a microwave (1000 W, Model 40GR47) in the air, equipped with firebricks for further insulation and microwaved for a total of 30 minutes at 5-minute intervals with grinding in between. The sample was then pressed into a pellet and placed in a furnace pre-heated to 1250 ˚C in air for 1 hr, then air quenched. The resulting sample is a homogeneous brown chunk that is stable in air. A high purity phase could not be realized without a 4% deficiency in $Ta_2O_5$ or microwaving, the reasons for which are detailed later.

**Phase Identification:**

($Mn_{0.2}Fe_{0.2}Co_{0.2}Ni_{0.2}Cu_{0.2}$)$Ta_{1.92}O_{6-\delta}$ was crushed into a powder and prepared for powder X-ray diffraction (XRD). A Bruker D2 PHASER was used with Cu $K_\alpha$ radiation ($\lambda$ = 1.54060 Å, Ge monochromator). The Bragg angle measured was from 5 to 100 ° at a rate of 1.7 °/min with a step of 0.012 °. Rietveld fitting in FULLPROF was employed to analyze the crystal structure and test the phase purity of ($Mn_{0.2}Fe_{0.2}Co_{0.2}Ni_{0.2}Cu_{0.2}$)$Ta_{1.92}O_{6-\delta}$.[24]

**Physical Properties Measurements:**

Magnetic Properties were measured in a Quantum Design Dynacool physical properties measurement system (PPMS) (1.8- 300 K, 0-9 T) equipped with ACMS II with both DC and AC magnetic fields available as well as a vibrating-sample magnetometer (VSM). Heat capacity was measured in the PPMS from 1.9 K to 37 K. Resistivity measurements were conducted in the PPMS from 310 K to 350 K using the four-probe method with platinum wires attached to the representative sample using silver epoxy.

**Scanning Electron Microscopy (SEM) with Energy-Dispersive Spectroscopy (EDS):**

Compositional analysis was performed via scanning electron microscopy (SEM) with energy-dispersive spectroscopy (EDS). A Ziess Sigma 500 VP SEM with Oxford Aztec X-EDS was used with an electron beam energy of 20 kV.

***Results and Discussion***

**Crystal structure and compositional analysis:** Powder XRD pattern of the air-quenched sample from 5 to 100 ° is shown in Figure 1(a). The pattern was fitted via Rietveld method by using the crystal structure of reported $CoTa_2O_6$[25]. The fitting parameters, i.e., $R_p$= 3.62 %, $R_{wp}$ = 4.63 %, $R_{exp}$ = 4.02 % and $\chi^2$ = 1.32, showed good match between the trirutile $CoTa_2O_6$ and the proposed HEO $(Mn_{0.2}Fe_{0.2}Co_{0.2}Ni_{0.2}Cu_{0.2})Ta_{1.92}O_{6-\delta}$. No prominent secondary phase was found, indicating that $(Mn_{0.2}Fe_{0.2}Co_{0.2}Ni_{0.2}Cu_{0.2})Ta_{1.92}O_{6-\delta}$ crystallizes in the same space group as $CoTa_2O_6$, i.e., $P\ 4_2/mnm$ with a high purity. The refined lattice parameters of $(Mn_{0.2}Fe_{0.2}Co_{0.2}Ni_{0.2}Cu_{0.2})Ta_{1.92}O_{6-\delta}$ are $a$ = 4.73179(1) Å , $c$ = 9.20395(3) Å , which creates a slightly smaller unit cell than $CoTa_2O_6$ ($a$ = 4.73580 Å , $c$ = 9.17080 Å) with smaller tetragonality, i.e., smaller c/a ratio. The fitted atomic coordinates are shown in Table S1 in the supporting information (SI), which further confirms the trirutile structure for $(Mn_{0.2}Fe_{0.2}Co_{0.2}Ni_{0.2}Cu_{0.2})Ta_{1.92}O_{6-\delta}$ . The obtained crystal structure is shown in Figures 1(b) & 1(c). Edge-shared $M@O_6$ and $Ta@O_6$ octahedra stack along the $c$ axis and are connected with each other in a corner-sharing manner. The $M@O_6$ layers with high-entropy site consisting of 3*d* transition metals within *ab* plane are separated by $Ta@O_6$ layers.

In order to confirm the high-entropy nature of $(Mn_{0.2}Fe_{0.2}Co_{0.2}Ni_{0.2}Cu_{0.2})Ta_{1.92}O_{6-\delta}$, SEM-EDS analysis was performed. In two distinct regions of $(Mn_{0.2}Fe_{0.2}Co_{0.2}Ni_{0.2}Cu_{0.2})Ta_{1.92}O_{6-\delta}$, elemental maps were collected and can be seen in Figure 1(d), for which the results are shown in Table S2 in the supporting information. The average chemical formula was found to be $Mn_{0.19(1)}Fe_{0.20(1)}Co_{0.18(1)}Ni_{0.20(1)}Cu_{0.20(2)}Ta_{1.92(1)}O_{6-\delta}$, displaying near equiatomic ratios of each 3*d* transition metal. Given that each 3*d* transition metal M can form a stable phase in $MTa_2O_6$[21,22,26,27], our material is not an entropy-stabilized material.[6] Meanwhile, we note that there is a deficiency in oxygen based on EDS and further examined in the magnetic superexchange pathway section. Since $(Mn_{0.2}Fe_{0.2}Co_{0.2}Ni_{0.2}Cu_{0.2})Ta_{1.92}O_{6-\delta}$ is an insulator (as discussed later), charging effects may affect the EDS results[28]. When considering this, our results are within the error of loading composition. To study the surface composition of our sample, XPS was performed. Three points of interest were analyzed to determine the surface concentration of the 3*d* transition metals to be

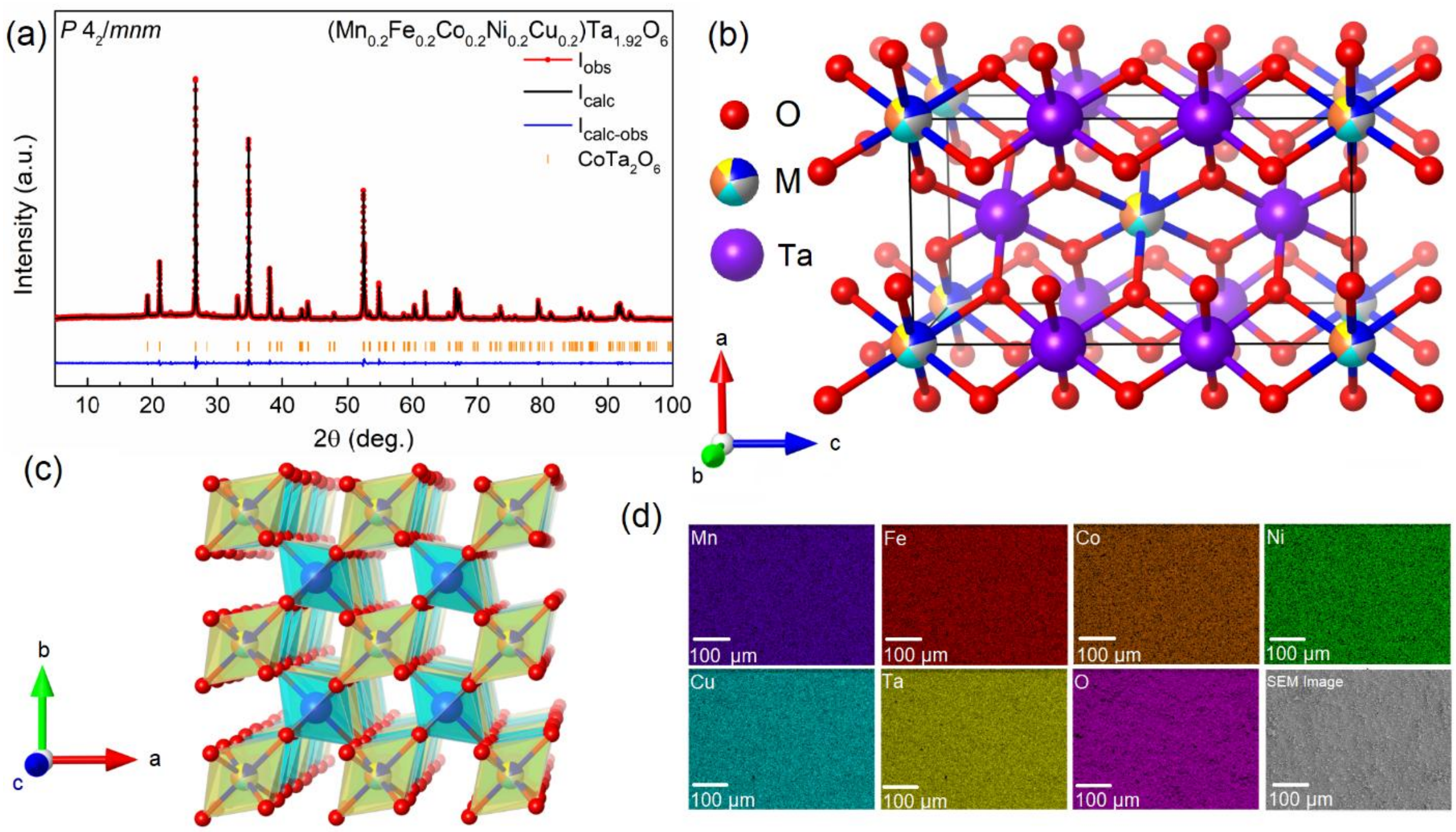

**Figure 1 (a)** Powder XRD pattern of $(Mn_{0.2}Fe_{0.2}Co_{0.2}Ni_{0.2}Cu_{0.2})Ta_{1.92}O_{6-\delta}$ . The observed pattern is shown by the red line with dots while the black line represents the calculated pattern. The difference between the observed and calculated pattern is shown in blue. Bragg peaks positions of $CoTa_2O_6$ are visualized with orange vertical bars. **(b)** $(Mn_{0.2}Fe_{0.2}Co_{0.2}Ni_{0.2}Cu_{0.2})Ta_{1.92}O_{6-\delta}$ trirutile structure viewing from the *b* axis. Here, "M" stands for the 3*d* transition metals in $(Mn_{0.2}Fe_{0.2}Co_{0.2}Ni_{0.2}Cu_{0.2})Ta_{1.92}O_{6-\delta}$. **(c)** $(Mn_{0.2}Fe_{0.2}Co_{0.2}Ni_{0.2}Cu_{0.2})Ta_{1.92}O_{6-\delta}$ viewing from the *c* axis with Ta@$O_6$ (cyan) and M@$O_6$ (yellow) octahedra **(d)** SEM-EDS elemental mapping results of Co, Cu, Fe, Mn, Ni, Ta, and O on a ~0.11-$mm^2$ area of pellet sample.

14(5)% Mn, 35(4)% Fe, 27(8)% Co, 10(2)% Ni, and 14(10)% Cu. As exemplified by the high levels of deviation as well as the non-equiatomic ratios, there are surface inhomogeneities in the sample. More analysis of XPS is provided in the SI. XPS has a penetration depth of ~10 nm[29], while EDS has a penetration depth of around 2 microns at 20keV.[30] Therefore, the surface inhomogeneities will less likely affect the bulk properties. Since our studies focus on the bulk properties of $(Mn_{0.2}Fe_{0.2}Co_{0.2}Ni_{0.2}Cu_{0.2})Ta_{1.92}O_{6-\delta}$, the EDS determined chemical composition will reflect more precisely the bulk composition. Therefore, all subsequent calculations were performed based on the loading composition, i.e., $(Mn_{0.2}Fe_{0.2}Co_{0.2}Ni_{0.2}Cu_{0.2})Ta_{1.92}O_{6-\delta}$. Further information about XPS results can be found in the Supporting Information.

**Short-range antiferromagnetic ordering in $(Mn_{0.2}Fe_{0.2}Co_{0.2}Ni_{0.2}Cu_{0.2})Ta_{1.92}O_{6-\delta}$:** $(Mn_{0.2}Fe_{0.2}Co_{0.2}Ni_{0.2}Cu_{0.2})Ta_{1.92}O_{6-\delta}$ was measured for the temperature dependence of magnetic susceptibility from 1.8 to 300 K under an external magnetic field of 0.1 T in a zero-field cooling mode, as shown in Figure 2. Paramagnetic behavior was seen under high temperature while a broad peak can be observed at ~ 4 K, as discussed later. Curie-Weiss (CW) fitting was performed from 150 to 275 K according to the modified CW law:

$$\chi = \frac{C}{T - \theta_{CW}} + \chi_0$$

where $\chi$ is the magnetic susceptibility, C is the Curie constant independent of temperature and related to the effective moment ($\mu_{eff}$), $\theta_{CW}$ is the CW temperature and $\chi_0$ is a constant independent of temperature and related to the core diamagnetism and temperature-independent paramagnetic contributions such as Pauli paramagnetism. The $\chi_0$ parameter was added to CW fitting since a slightly positive curvature was noted at ~ 250K and was fitted to be 0.00109(1) emu/Oe/mol. The resulted $\theta_{CW}$ is -19.6 (2) K, indicating antiferromagnetic (AFM) spin-spin coupling within the fitted temperature region. Therefore, the slight deviation from CW behavior at ~ 58 K might originate from the onset of AFM coupling, as seen in Figure S1 in the SI. The resulting $\theta_{CW}$ is enhanced in comparison to the parent Mn (-4.2 K)[26], Fe (-8.11(5) K)[27], Co (-6.63(5) K)[22], Ni (-10.3(2) K)[22] and Cu (-1.5(2) K)[21] tantalate compounds.

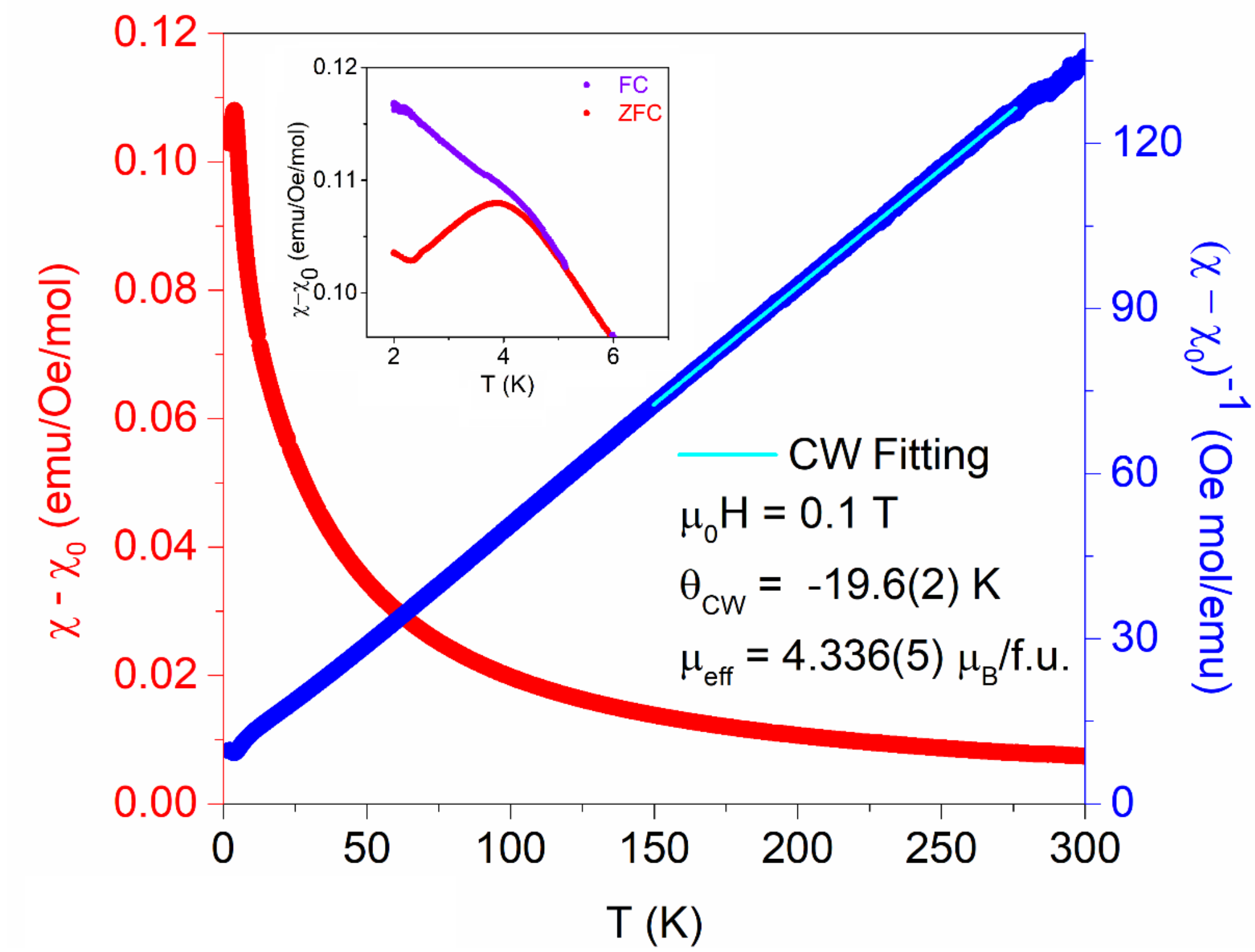

**Figure 2** Magnetic susceptibility minus $\chi_0$ (left red axis) and inverse of magnetic susceptibility minus $\chi_0$ (right blue axis) of $(Mn_{0.2}Fe_{0.2}Co_{0.2}Ni_{0.2}Cu_{0.2})Ta_{1.92}O_{6-\delta}$ measured under an external magnetic field of 0.1 T with respect to temperature. The cyan line represents the Curie-Weiss fitting. The inset zooms in on the difference between the inverse of magnetic susceptibility minus $\chi_0$ with respect to temperature under zero-field cooling (red line) and field cooling (purple line) also at an external magnetic field of 0.1 T.

Based on $\mu_{eff}$ (spin-only) = $\sqrt{n(n+2)}$ where $n$ is the number of unpaired electrons, the spin-only moment for all 3$d$ transition metal ions except for $Co^{2+}$ in high-spin octahedral crystal field in our material are as follows: $Mn^{2+}$ (~ 5.92 $\mu_B$), $Fe^{3+}$ (~ 5.92 $\mu_B$), $Ni^{2+}$ (~ 2.83 $\mu_B$), and $Cu^{2+}$ (~ 1.73 $\mu_B$). Since $Co^{2+}$ ions exhibit large single-ion anisotropy, we chose to use the effective moment reported for $CoTa_2O_6$, 4.9(2) $\mu_B$ for $Co^{2+}$ instead.[23] The $\mu_{eff}$ when considering equal contributions from each 3d transition metal, averages to ~ 4.26 $\mu_B$ which is consistent with the $\mu_{eff}$ from CW fitting determined via $\mu_{eff} = \sqrt{8C}$ to be 4.336 (5) $\mu_B$/f.u. The larger fitted $\mu_{eff}$ might be due to the unaccounted contributions from orbital angular momentum.

As shown in Figure 2 inset, the $\chi$ vs T curve features a broad downward parabola shape at ~ 4 K when under zero-field cooling (ZFC), unlike the more linear shape that is realized under field-cooling (FC). Therefore, we performed the temperature-dependent AC magnetic susceptibility measurements with a DC field of 100 Oe and an AC field of 10 Oe under different frequencies to confirm the existence of spin-glass state. As shown in Figure S2 in the SI, no frequency-dependent behavior can be seen for the AC susceptibility under 936 Hz, 1635 Hz and 1858 Hz from 2 K to 10 K, which excludes conventional spin-glass state in $(Mn_{0.2}Fe_{0.2}Co_{0.2}Ni_{0.2}Cu_{0.2})Ta_{1.92}O_{6-\delta}$. Meanwhile, the DC magnetic susceptibility under FC protocol does not reach a plateau in our material, which further confirms the lack of conventional spin-glass state. The divergence between the ZFC and FC curves below ~4 K normally originates from spin freezing.[31,32] Therefore, the broad peak observed in Figure 2 might originate from short range antiferromagnetic ordering, which will be further discussed later along with the heat capacity results. Moreover, similar behavior can be seen in Mg-doped $CoTa_2O_6$ where short-range magnetic ordering was observed when the Mg concentration is above ~10%.[23]

The field-dependence of magnetization, as shown in Figure 3, illustrates that at 2, 10, and 300 K, $(Mn_{0.2}Fe_{0.2}Co_{0.2}Ni_{0.2}Cu_{0.2})Ta_{1.92}O_{6-\delta}$ does not become saturated from up to 9 T. A slightly bent feature can be seen under 2 K, which is typical for paramagnetic state. In addition, no coercive field is observed, as can be seen in Figure S3 in the SI, which can be the indicative of the absence of a long-range ordered ferromagnetic component.

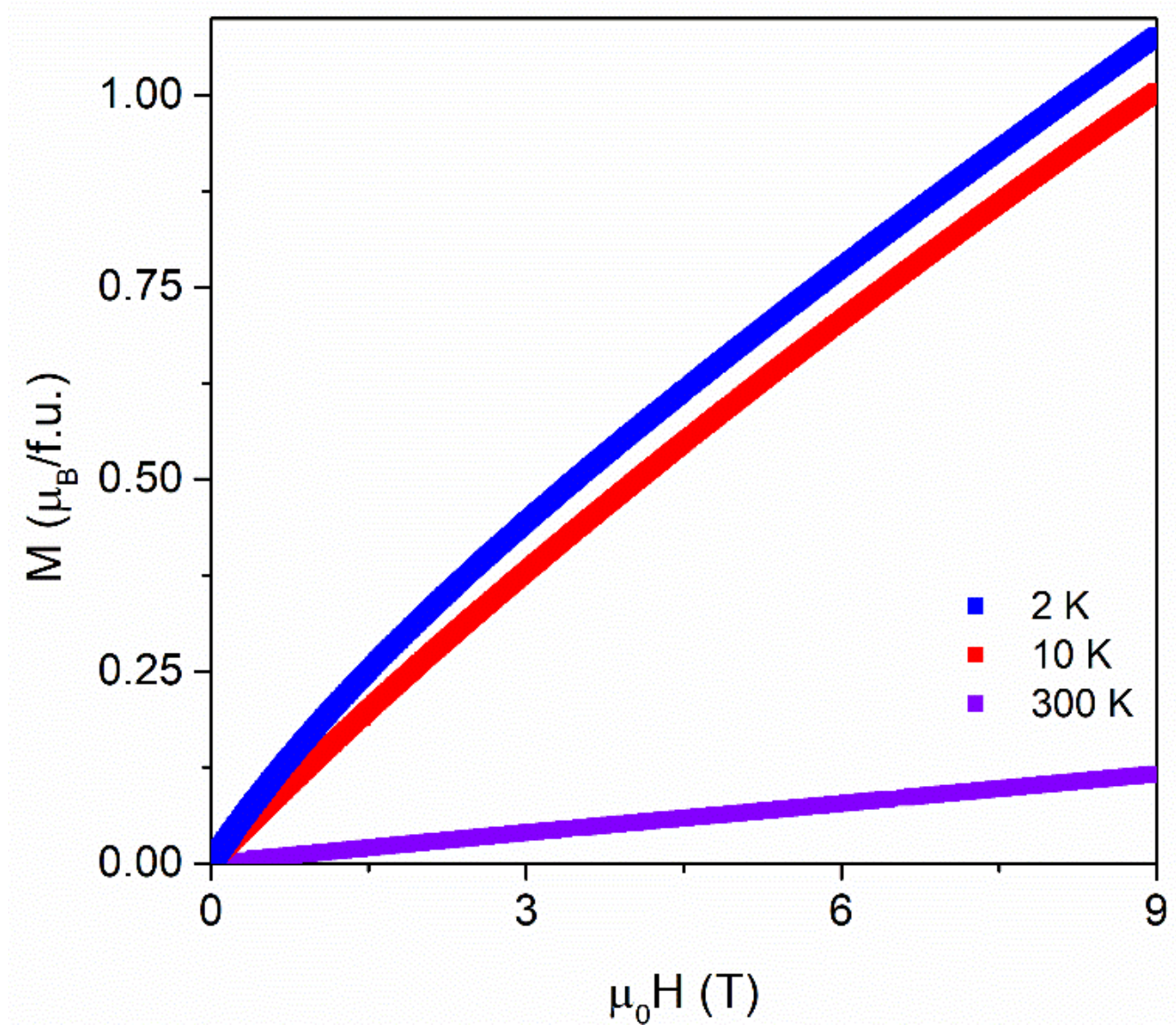

**Figure 3.** Field-dependent magnetization from 0-9 T at 2, 10, and 300 K, in blue, red and violet, respectively.

**Heat Capacity:** The temperature-dependent heat capacity of $(Mn_{0.2}Fe_{0.2}Co_{0.2}Ni_{0.2}Cu_{0.2})Ta_{1.92}O_{6-\delta}$ was measured from 2 to 35 K. No indication of long-range magnetic ordering, i.e., a λ-anomaly in heat capacity curve, can be seen, as shown in Figure 4(a). While a broad feature appears at ~4 K in Figure 4(a), Figure 4(b) plots the $C_{total}/T$ vs T curve and the broad peak is shown at ~ 4 K. The total heat capacity of a magnetic material can be treated as the sum of electronic, phononic, and magnonic contribution where $C_{total} = C_{el} + C_{ph} + C_{mag}$. As shown by resistivity in Figure S4 in the supporting information, $(Mn_{0.2}Fe_{0.2}Co_{0.2}Ni_{0.2}Cu_{0.2})Ta_{1.92}O_{6-\delta}$ exhibits insulating behavior, thus there are no electronic contributions to heat capacity, i.e., $C_{total} = C_{ph} + C_{mag}$. As an estimate, $C_{ph} = \beta T^3$ according to the Debye formula where $\beta$ is the vibrational contribution coefficient. However, just a $T^3$ term could not lead to a good fitting. This could be due to unaccounted high-frequency phononic contributions to heat capacity, as the Debye formula is only an estimate. Herein, $T^5$, $T^7$, $T^9$, and $T^{11}$ terms were added to obtain a reasonable fitting so that a polynomial fit was employed: $C_{ph}/T = \beta_1 T^2 + \beta_2 T^4 + \beta_3 T^6 + \beta_4 T^8 + \beta_5 T^{10}$where $\beta_1$, $\beta_2$, $\beta_3$, $\beta_4$, and $\beta_5$ are constants and were fitted to 0.00311(9) J/mol/K$^4$, -8.6(5) ×10$^{-6}$ J/mol/K$^6$, 1.2(1)×10$^{-8}$ J/mol/K$^8$, -7.7(8)×10$^{-12}$ J/mol/K$^{10}$, and 2.0(2)×10$^{-15}$ J/mol/K$^{12}$. Therefore, by subtracting the phononic contribution from $C_{total}/T$, pure $C_{mag}/T$ is plotted in Figure 4(b) as blue open circles. By integrating $C_{mag}/T$ vs T curve, the change

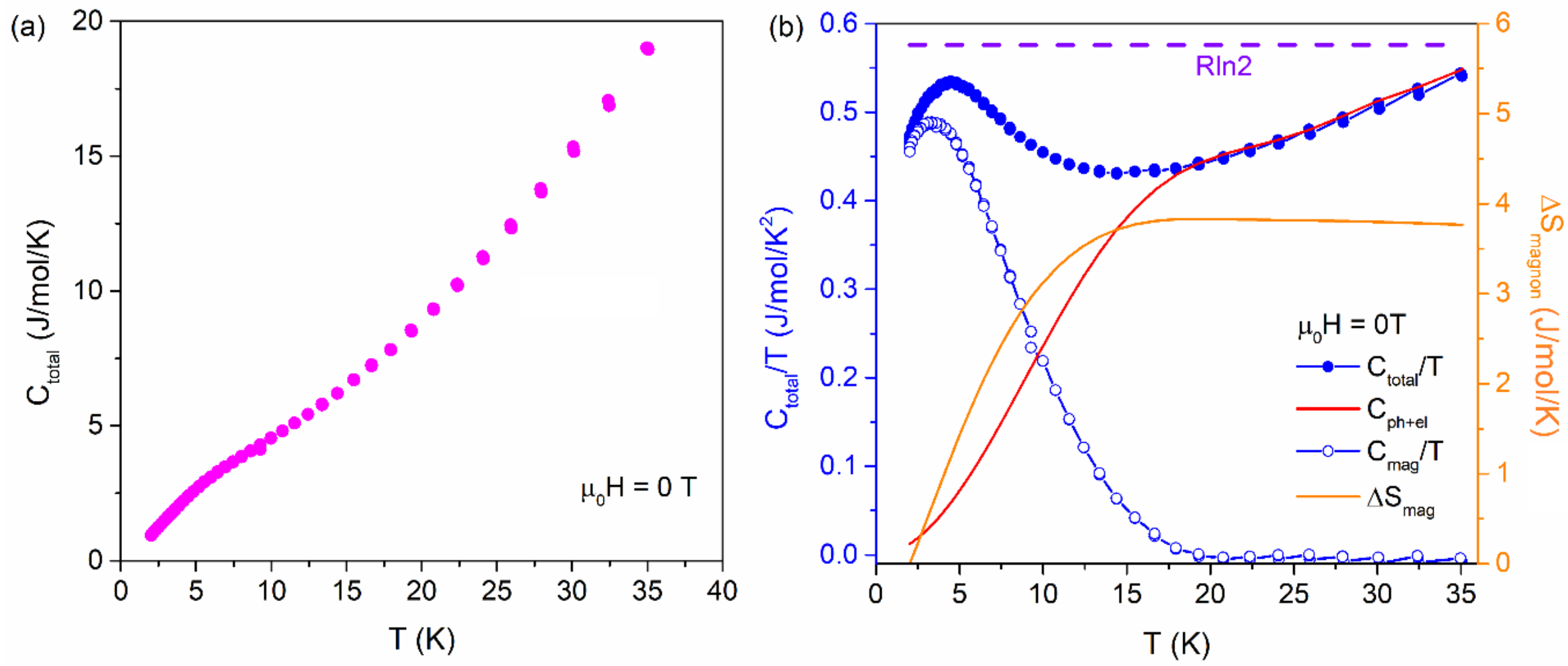

**Figure 4 (a).** $C_{total}$ is shown in magenta as a function of temperature. **(b)** $C_{total}/T$ is shown in blue filled circles while $C_{mag}/T$ is shown in blue open circles. The fitted $C_{ph}$ is graphed in red. These three curves correspond to the left blue axis. Rln2 is shown by the orange dashed line and $\Delta S_{mag}$ is the solid orange line; both lines correspond to the right orange axis.

in magnetic entropy, $\Delta S_{mag}$, is obtained. $\Delta S_{mag}$ is usually determined by the spin multiplicity in a magnetic system where contributions from orbital angular momentum can be ignored by $\Delta S_{mag}$ = Rln(2S+1) where R is the gas constant. Here, $\Delta S_{mag}$ does not exceed Rln2, which is the magnetic entropy change for a $S = ½$ system. Because the average spin state of the 3*d* transition metals appearing in $(Mn_{0.2}Fe_{0.2}Co_{0.2}Ni_{0.2}Cu_{0.2})Ta_{1.92}O_{6-\delta}$, i.e., $Mn^{2+}$ (S = 5/2), $Fe^{3+}$ (S = 5/2), $Co^{2+}$ (S = 3/2), $Ni^{2+}$ (S = 1), and $Cu^{2+}$ (S = 1/2), must be larger than $S = 3/2$. Therefore, the observed $\Delta S_{mag}$ is obviously smaller than Rln4, indicating the absence of long-range magnetic ordering within the measured temperature range in $(Mn_{0.2}Fe_{0.2}Co_{0.2}Ni_{0.2}Cu_{0.2})Ta_{1.92}O_{6-\delta}$. Heat capacity of $(Mn_{0.2}Fe_{0.2}Co_{0.2}Ni_{0.2}Cu_{0.2})Ta_{1.92}O_{6-\delta}$ was also measured under an external magnetic field, as shown in Figure S5 in the supporting information; no shift is observed for peak position. The field-independent behavior suggests that such anomaly below ~ 5 K is not spin-related, instead, it can be due to the phononic excitation.

**Magnetic superexchange interaction and possible origin of short-range magnetic ordering:** Based on the crystal structure of $(Mn_{0.2}Fe_{0.2}Co_{0.2}Ni_{0.2}Cu_{0.2})Ta_{1.92}O_{6-\delta}$, two possible magnetic superexchange pathways can be proposed, as illustrated in Figure 5, while the other pathways are related to them by symmetries. These superexchange pathways are based on the Goodenough-Kanamori rule[33,34] and the atomic orbitals of $M^{2+/3+}$ near the Fermi energy are determined by the semi-empirical extended-Hückel-tight-binding (EHTB) methods and CAESAR packages, as described in the SI. The calculation was only conducted on the parent compound, $CoTa_2O_6$, due to the complexities of the high-entropy $(Mn_{0.2}Fe_{0.2}Co_{0.2}Ni_{0.2}Cu_{0.2})Ta_{1.92}O_{6-\delta}$. However, the symmetries of molecular orbitals in $CoTa_2O_6$ can be well extended to $(Mn_{0.2}Fe_{0.2}Co_{0.2}Ni_{0.2}Cu_{0.2})Ta_{1.92}O_{6-\delta}$ regardless of addition or removal of valence electrons of 3*d* transition metal ions. Therefore, the analysis of molecular orbitals in the high-entropy compound can be conducted based on the highest-occupied molecular orbital in $CoTa_2O_6$ by varying the number of electrons.

The first superexchange pathway is shown in Figure 5(a) where two $M@O_6$ octahedra are connected by $Ta@O_6$ octahedron via an axial manner. In this case, half-filled $d_{xz}$ orbital of $M^{2+/3+}$, fully-filled $p_x$ and $p_z$ orbitals of $O^{2-}$ and empty $d_{z^2}$ orbital of $Ta^{5+}$ are present. The two $M^{2+/3+}$ ions on the opposite sides show antiferromagnetic coupling due to the following reasons: 1. due to the virtual electron hopping represented by the arrows, a pair of spins with opposite spin directions from oxygen $p_z$ orbitals can populate the empty $d_{z^2}$ orbital of $Ta^{5+}$, which provides a more stable ground state; 2. the exchange energies between the spins with the same direction on $p_x$ and $p_z$ orbitals of $O^{2-}$ stabilize the system as well, as indicated by the orange dashed lines in Figure 5; 3. virtual electron hopping can also occur between the $p_x$ orbitals of $O^{2-}$ and the half-filled $d_{xz}$ orbital of $M^{2+/3+}$, which leads to a stable singlet state. The superexchange interaction depicted results in the antiferromagnetic coupling between the $M^{2+/3+}$ ions. Similar reasons and superexchange pattern can be found in the second pathway, shown in Figure 5(b), while the involved atomic orbitals are now the half-filled $d_z{}^2$ orbital of $M^{2+/3+}$, fully-filled $p_x$ and $p_y$ orbitals of $O^{2-}$ and empty $d_{xy}$ orbital of $Ta^{5+}$. In both cases, antiferromagnetic coupling between M ions is dominant.

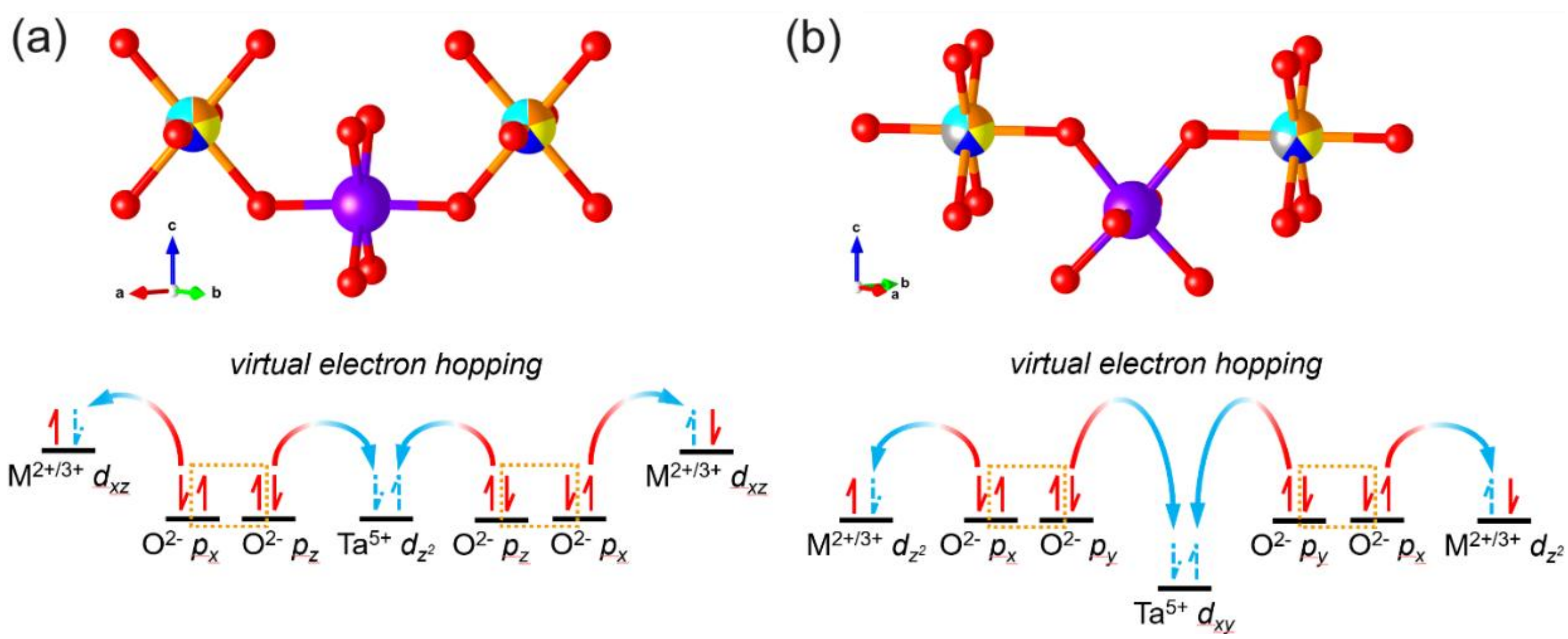

**Figure 5.** The two different magnetic superexchange pathways in $(Mn_{0.2}Fe_{0.2}Co_{0.2}Ni_{0.2}Cu_{0.2})Ta_{1.92}O_{6-\delta}$ between transition metal ions. $M^{2+/3+}$ stand for transition metal ions. The relative vertical distance between atomic orbitals stands for the relative geometry of ions. The dashed orange lines involve the pair of electrons on difference p orbitals of oxygen ions that stabilize the superexchange pathways by inclusion of exchange energy.

With the analysis above, we can speculate three possible reasons of the short-range magnetic ordering observed in magnetic properties measurements:

- The existence of $Ni^{2+}$ and $Cu^{2+}$ leads to the lack of half-filled $d_{xz}$ orbitals, which disables the scenario in Figure 5(a) where a half-filled $d_{xz}$ orbital of $M^{2+/3+}$ is necessary. Subsequently, the corresponding superexchange pathway is not functional due to the lack of electron hopping and, thus, magnetic coupling is absent. Meanwhile, all the $M^{2+/3+}$ ions possess a half-filled $d_z{}^2$ orbital, which allows the superexchange interaction in Figure 5(b). Therefore, with part of the superexchange pathways disabled, long-range magnetic ordering does not exist in $(Mn_{0.2}Fe_{0.2}Co_{0.2}Ni_{0.2}Cu_{0.2})Ta_{1.92}O_{6-\delta}$.
- Due to the existence of $Fe^{3+}$, in order to maintain charge neutrality, possible removal of cations can take place as a result of ionic charge compensation mechanism. Therefore, part of the $M^{2+}$ or $Ta^{5+}$ sites may be vacant, leading to the interruption of the superexchange pathways and the absence of long-range magnetic ordering. The $Ta^{5+}$ vacancy is seen in the inability to synthesis a high purity sample with non-deficient $Ta_2O_5$ amounts. In addition, to maintain charge neutrality, there must be oxygen deficiencies that similarly disrupt the superexchange pathway.
- The high-entropy nature of the $M^{2+/3+}$ site can result in the local distortion of the $M@O_6$ octahedra, which reduces the orbital overlap between $M^{2+/3+}$ and the intermediate $O^{2-}$. Consequently, electron hopping may not occur due to the altered symmetries of orbitals.

**Conclusion**

In this paper, we reported the first high-entropy oxide in a trirutile structure. High-purity phase with high configurational disorder in a uniform structure has been determined via powder XRD and EDS measurements. The new material, $(Mn_{0.2}Fe_{0.2}Co_{0.2}Ni_{0.2}Cu_{0.2})Ta_{1.92}O_{6-\delta}$, shows antiferromagnetic spin-spin coupling under high temperatures while exhibiting short-range antiferromagnetic ordering under low temperatures. Heat capacity further shows the evidence of the absence of long-range magnetic ordering in this compound. XPS shows the presence of divalent Mn, Co, Ni and Cu, along with trivalent Fe, which are consistent with the observation in the magnetic results. The discovery of the first trirutile high-entropy oxide builds a new platform for investigating the interplay between high-entropy nature and their magnetism. Moreover, it allows the manipulation of a wider range of physical properties in high-entropy oxides.

## *Associated Content*

Supporting Information

The Supporting Information is available free of charge at XXX.

Deviation from Curie-Weiss behavior below ~50 K. AC magnetic susceptibility data. Magnetic hysteresis loop. Temperature-dependent resistivity data. Heat capacity data under zero and applied magnetic field. Powder XRD fitted atomic coordinates. EDS results.

## *Author Information*

Corresponding Author: xig75@pitt.edu

Notes: The authors declare no competing financial interest.

## *Acknowledgements*

G. A. and X.G. are supported by the startup fund from the University of Pittsburgh and the Pitt Momentum Fund. Portions of this work were performed at the Molecular Foundry and supported by the Office of Science, Office of Basic Energy Sciences, of the U.S. Department of Energy under Contract No. DE-AC02-05CH11231. Work at the University of Arizona is supported by the National Science Foundation under Award No. DMR-2338229.

**For Table of Contents Only**

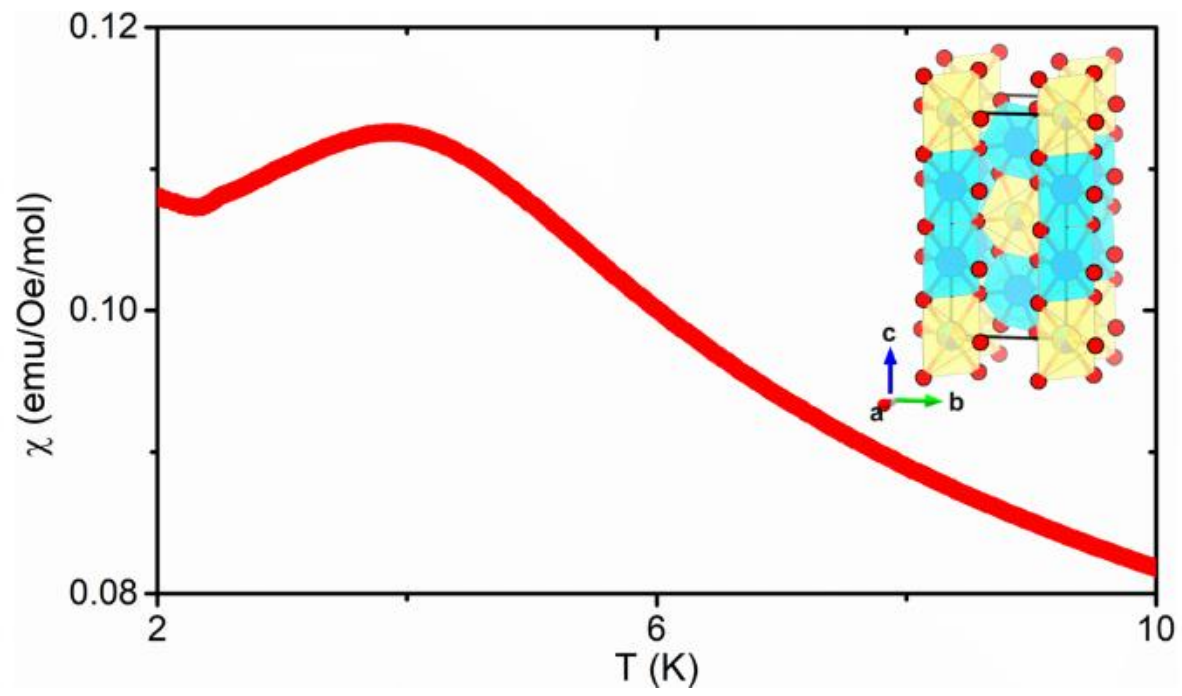

***Supporting Information***

# Absence of Long-Range Magnetic Ordering in a Trirutile High-Entropy Oxide $(Mn_{0.2}Fe_{0.2}Co_{0.2}Ni_{0.2}Cu_{0.2})Ta_{1.92}O_{6-\delta}$

*Gina Angelo,[a] Liana Klivansky,[b] Jeremy G. Philbrick,[c] Tai Kong,[c,d] Jian Zhang,[b] Xin Gui [a*]*

[a] Department of Chemistry, University of Pittsburgh, Pittsburgh, PA, 15260, USA
[b] The Molecular Foundry, Lawrence Berkeley National Laboratory, Berkeley, CA, 94720, USA
[c] Department of Physics, University of Arizona, Tucson, AZ, 85721, USA
[d] Department of Chemistry and Biochemistry, University of Arizona, Tucson, AZ, 85721, USA

## Table of Contents

**Table S1.** The atomic sites of $(Mn_{0.2}Fe_{0.2}Co_{0.2}Ni_{0.2}Cu_{0.2})Ta_{1.92}O_{6-\delta}$ were determined via Rietveld refinement. $(Mn_{0.2}Fe_{0.2}Co_{0.2}Ni_{0.2}Cu_{0.2})Ta_{1.92}O_{6-\delta}$ is in space group $P4_2/mnm$ (no. 136) with parameters $a$ = 4.73179 (1) Å and $c$ = 9.20395 (3) Å. M stands for 3*d* transition metals.

| | ***x*** | ***y*** | ***z*** |
|---|---|---|---|
| **M1** | 0 | 0 | 0 |
| **Ta1** | 0 | 0 | 0.3317 (1) |
| **O1** | 0.2896 (9) | 0.2896 (9) | 0 |
| **O2** | 0.3007 (6) | 0.3007 (6) | 0.3337 (6) |

**Table S2** EDS results of $MTa_{1.92}O_6$ for two different areas of the sample. Data was normalized to Ta and compared to the loading composition.

| Spectra | Mn | Fe | Co | Ni | Cu | Ta |
|---|---|---|---|---|---|---|
| Map 1 | 2.18 | 2.33 | 2.13 | 2.27 | 2.46 | 22.07 |
| Map 2 | 2.30 | 2.31 | 2.09 | 2.33 | 2.16 | 22.18 |
| Average | 2.24 (8) | 2.32 (2) | 2.11 (2) | 2.30 (4) | 2.31 (21) | 22.13 (7) |
| Normalized to Ta as 1.92 | 0.19 (1) | 0.20 (1) | 0.18 (1) | 0.20 (1) | 0.20 (2) | 1.92 (1) |
| Loading Composition | 0.2 | 0.2 | 0.2 | 0.2 | 0.2 | 1.92 |

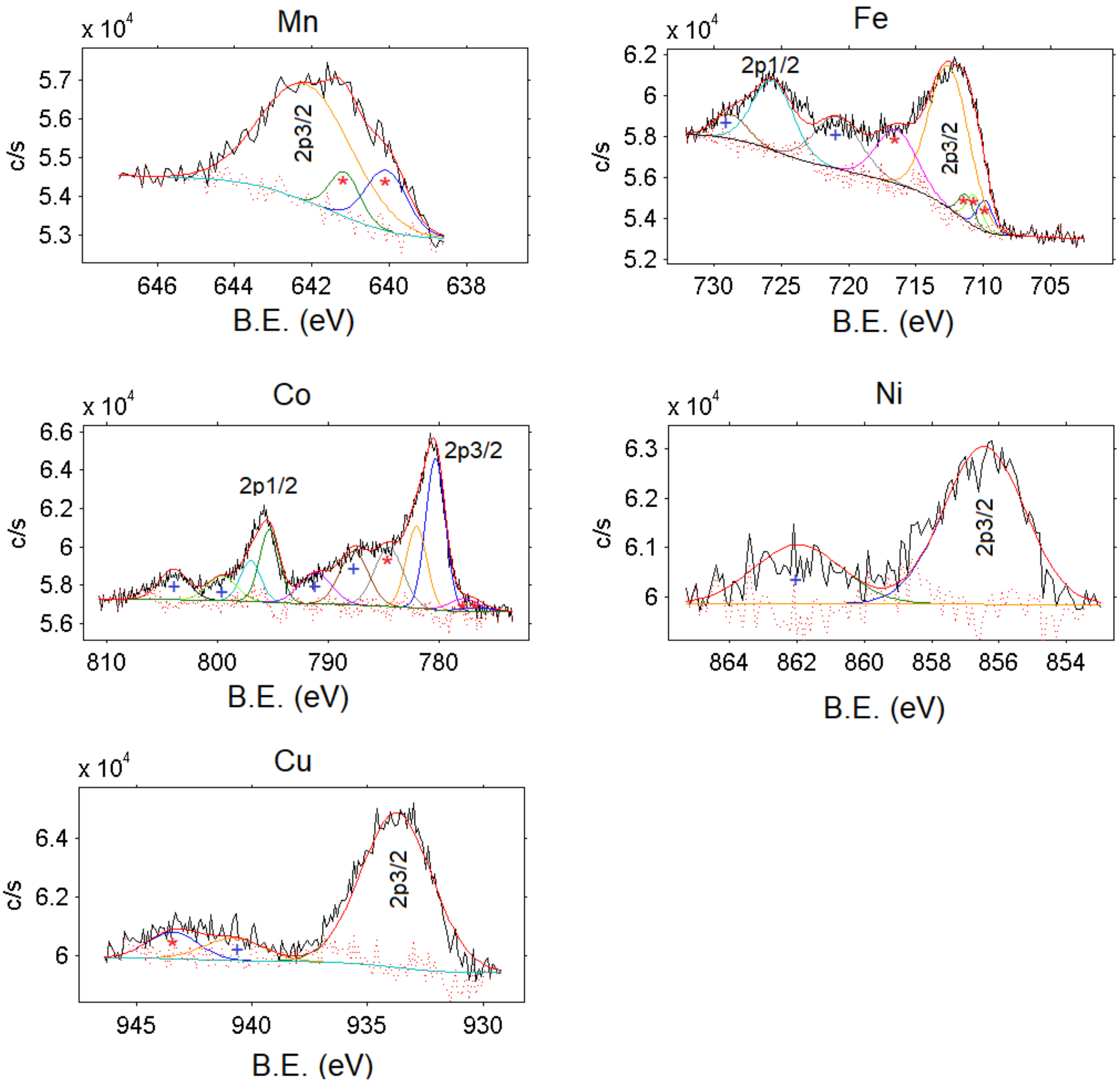

**Figure S3**. Peak fitting of Mn, Fe, Co, Ni, and Cu in $(Mn_{0.2}Fe_{0.2}Co_{0.2}Ni_{0.2}Cu_{0.2})Ta_{1.92}O_{6-\delta}$ in area 1. B.E. stands for binding energy. Blue plus signs mark satellite peaks wile red asterisks mark Auger peaks.

**Table S3.** Peak fitting of Mn, Fe, Co, Ni, and Cu regions in $(Mn_{0.2}Fe_{0.2}Co_{0.2}Ni_{0.2}Cu_{0.2})Ta_{1.92}O_{6-\delta}$ in area 1.

| Element/ Transition | Peak Energy (eV) | Peak FWHM (eV) | Peak Area (eV*cts/s) | Peak Assignment | RSF | Atomic % |
|---|---|---|---|---|---|---|
| Cu LMM | 640.07 | 1.27 | 2124 | Cu Auger | | |
| Ni LMM | 641.16 | 0.92 | 1145 | Ni Auger | | |
| Mn 2p3/2 | 642.14 | 2.74 | 8835 | $Mn^{2+}$ | 1.757 | 14.38 |
| Ni LMM | 709.85 | 1.40 | 2254 | Ni Auger | | |
| Cu LMM | 710.68 | 1.40 | 2254 | Cu Auger | | |
| Co LMM | 711.28 | 1.31 | 1793 | Co Auger | | |
| Fe 2p3/2 | 712.53 | 3.49 | 25885 | $Fe^{3+}$ | 2.946 | 37.58 |
| Cu LMM | 716.45 | 3.52 | 9841 | Cu Auger | | |
| Fe 2p Sat | 720.82 | 4.05 | 10933 | Satellite | | |
| Fe 2p1/2 | 725.63 | 3.57 | 12943 | $Fe^{3+}$ | | |
| Fe2p sat | 728.82 | 2.88 | 3684 | Satellite | | |
| Co LMM | 776.58 | 3.58 | 758 | Co Auger | | |
| Ni LMM | 777.79 | 3.08 | 2177 | Ni Auger | | |
| Co 2p3/2 | 780.32 | 2.08 | 17520 | $Co^{2+}$ | 3.529 | 12.04 |
| Co 2p3/2 | 782.04 | 2.20 | 9975 | $Co^{3+}$ | 3.529 | 20.41 |
| Fe LMM | 784.55 | 3.08 | 10108 | Fe Auger | | |
| Co 2p sat | 787.67 | 3.08 | 9230 | Satellite | | |
| Co 2p sat | 791.16 | 3.46 | 5907 | Satellite | | |
| Co 2p1/2 | 795.29 | 2.16 | 8760 | $Co^{2+}$ | | |
| Co2p1/2 | 797.01 | 2.20 | 4988 | $Co^{3+}$ | | |
| Co 2p sat | 799.60 | 3.08 | 4216 | Satellite | | |
| Co 2p sat | 803.96 | 3.08 | 5244 | Satellite | | |
| Ni 2p3/2 | 856.45 | 2.82 | 9621 | $Ni^{2+}$ | 2.309 | 12.30 |
| Ni 2p3 sat | 861.97 | 3.04 | 3861 | Satellite | | |
| Cu 2p3/2 | 933.73 | 3.26 | 20323 | $Cu^{2+}$ | 2.626 | 3.29 |
| Cu 2p sat | 940.79 | 3.06 | 2462 | Satellite | | |
| Mn LMM | 943.43 | 2.56 | 2474 | Mn Auger | | |

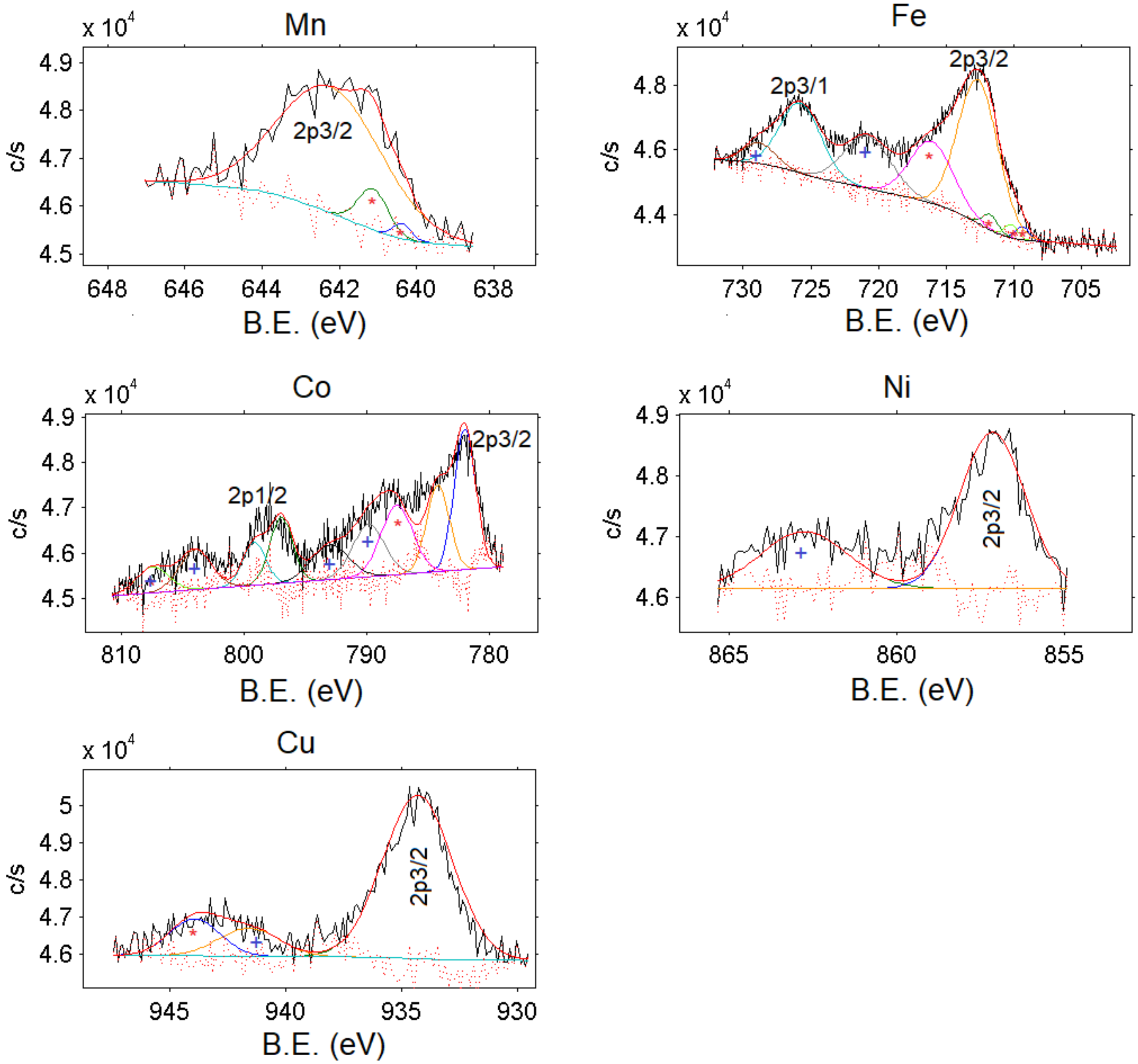

**Figure S4**. Peak fitting of Mn, Fe, Co, Ni, and Cu in $(Mn_{0.2}Fe_{0.2}Co_{0.2}Ni_{0.2}Cu_{0.2})Ta_{1.92}O_{6-\delta}$ in area 2. B.E. stands for binding energy. Blue plus signs mark satellite peaks wile red asterisks mark Auger peaks.

**Table S4**. Peak fitting of Mn, Fe, Co, Ni, and Cu regions in $(Mn_{0.2}Fe_{0.2}Co_{0.2}Ni_{0.2}Cu_{0.2})Ta_{1.92}O_{6-\delta}$ in area 2.

| Element/ Transition | Peak Energy (eV) | Peak FWHM (eV) | Peak Area (eV*cts/s) | Peak Assignment | RSF | Atomic % |
|---|---|---|---|---|---|---|
| Cu LMM | 640.38 | 0.56 | 193 | Cu Auger | | |
| Ni LMM | 641.13 | 0.91 | 828 | Ni Auger | | |
| Mn 2p3/2 | 642.23 | 3.20 | 8980 | $Mn^{2+}$ | 1.757 | 18.89 |
| Ni LMM | 709.37 | 1.15 | 403 | Ni Auger | | |
| Cu LMM | 710.2 | 1.15 | 492 | Cu Auger | | |
| Co LMM | 711.75 | 1.31 | 677 | Co Auger | | |
| Fe 2p3/2 | 712.66 | 3.49 | 16372 | $Fe^{3+}$ | 2.946 | 30.71 |
| Cu LMM | 716.10 | 3.78 | 7633 | Cu Auger | | |
| Fe 2p3Sat | 720.82 | 4.05 | 7005 | Satellite | | |
| Fe 2p1/2 | 725.76 | 3.57 | 8186 | $Fe^{3+}$ | | |
| Fe2p sat | 728.82 | 2.78 | 1859 | Satellite | | |
| Co 2p3/2 | 781.98 | 2.07 | 6827 | $Co^{2+}$ | 3.529 | 10.65 |
| Co 2p3/2 | 784.18 | 2.20 | 4455 | $Co^{3+}$ | 3.529 | 6.95 |
| Fe LMM | 787.53 | 3.08 | 4993 | Fe Auger | | |
| Co 2p sat | 789.83 | 3.08 | 3788 | Satellite | | |
| Co 2p sat | 792.98 | 3.58 | 2909 | Satellite | | |
| Co 2p1/2 | 796.95 | 2.20 | 3413 | Co3O4 | | |
| Co2p1/2 | 799.15 | 2.20 | 2228 | Co3O4 | | |
| Co 2p sat | 803.94 | 3.08 | 2904 | Satellite | | |
| Co 2p sat | 807.43 | 3.08 | 1807 | Satellite | | |
| Ni 2p3 | 857.15 | 2.27 | 6231 | $Ni^{2+}$ | 2.309 | 9.85 |
| Ni 2p3 sat | 862.76 | 3.04 | 3030 | Satellite | | |
| Cu 2p3/2 | 934.30 | 3.56 | 16632 | $Cu^{2+}$ | 2.626 | 22.96 |
| Cu 2p sat | 941.59 | 3.06 | 2453 | Satellite | | |
| Mn LMM | 943.94 | 2.56 | 2712 | Mn Auger | | |

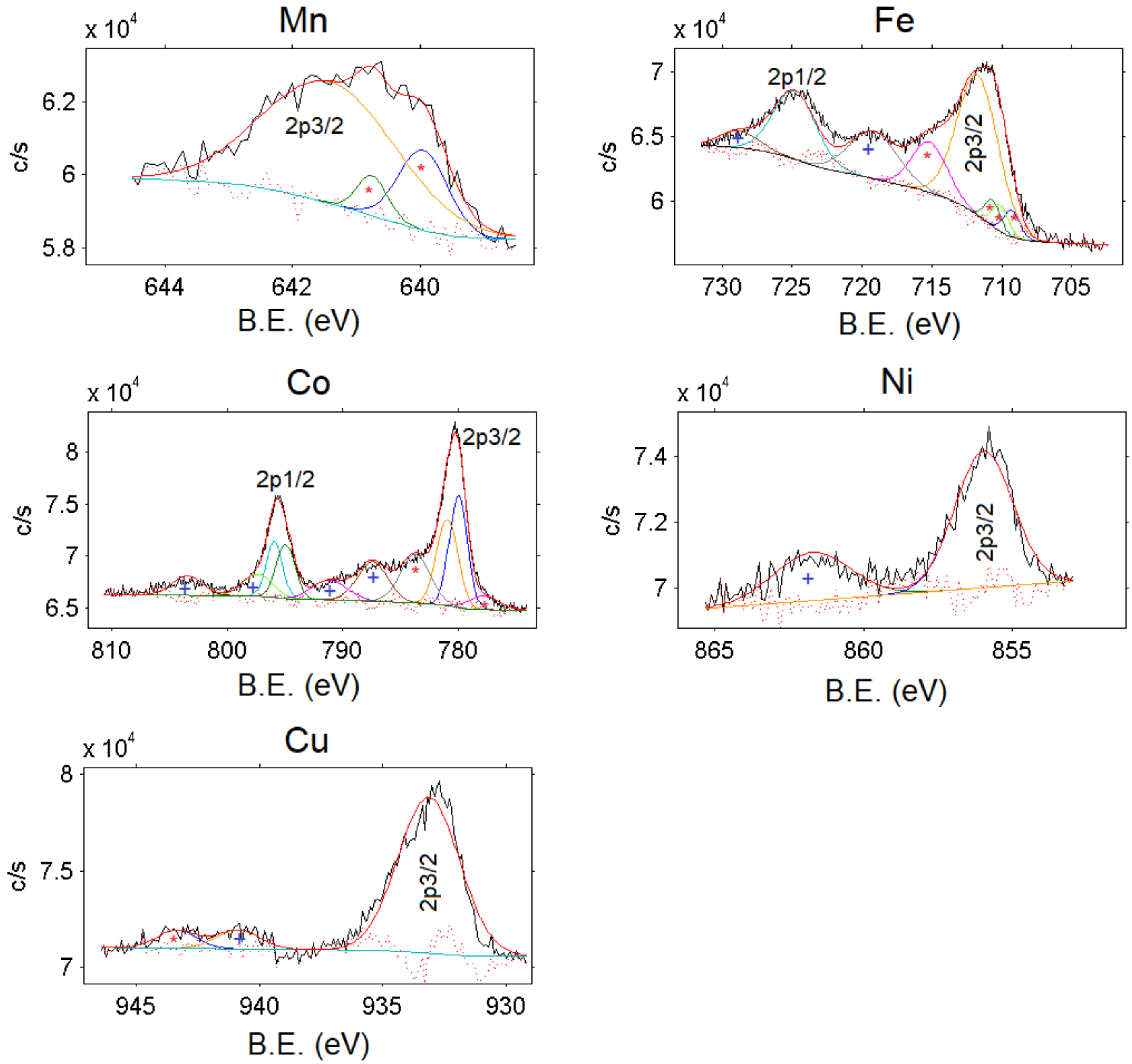

**Figure S5.** Peak fitting of Mn, Fe, Co, Ni, and Cu in $(Mn_{0.2}Fe_{0.2}Co_{0.2}Ni_{0.2}Cu_{0.2})Ta_{1.92}O_{6-\delta}$ in area 3. B.E. stands for binding energy. Blue plus signs mark satellite peaks wile red asterisks mark Auger peaks.

**Table S5**. Peak fitting of Mn, Fe, Co, Ni, and Cu regions in $(Mn_{0.2}Fe_{0.2}Co_{0.2}Ni_{0.2}Cu_{0.2})Ta_{1.92}O_{6-\delta}$ in area 3.

| Element/ Transition | Peak Energy (eV) | Peak FWHM (eV) | Peak Area (eV*cts/s) | Peak Assignment | RSF | Atomic (%) |
|---|---|---|---|---|---|---|
| Cu LMM | 639.95 | 0.91 | 2134 | Cu Auger | | |
| Ni LMM | 640.75 | 0.56 | 659 | Satellite | | |
| Mn 2p3/2 | 641.42 | 2.49 | 8822 | Mn Auger | | |
| Cu LMM | 639.95 | 0.91 | 2134 | Cu Auger | | |
| Ni LMM | 640.75 | 0.56 | 659 | Ni Auger | | |
| Mn 2p3/2 | 641.42 | 2.49 | 8822 | $Mn^{2+}$ | 1.757 | 8.90 |
| Ni LMM | 709.29 | 1.4 | 3024 | Ni Auger | | |
| Cu LMM | 710.12 | 1.4 | 3024 | Cu Auger | | |
| Co LMM | 710.67 | 1.31 | 2832 | Co Auger | | |
| Fe 2p3/2 | 711.72 | 3.49 | 40272 | $Fe^{3+}$ | 2.946 | 36.23 |
| Cu LMM | 714.97 | 3.02 | 11908 | Cu Auger | | |
| Fe 2p Sat | 719.18 | 4.05 | 15814 | Satellite | | |
| Fe 2p1/2 | 724.82 | 3.57 | 20136 | $Fe^{3+}$ | | |
| Fe2p sat | 728.82 | 2.78 | 1859 | Satellite | | |
| Ni LMM | 777.79 | 3.08 | 4287 | Ni Auger | | |
| Co 2p3/2 | 779.98 | 1.85 | 21280 | $Co^{2+}$ | 3.529 | 15.92 |
| Co 2p3/2 | 780.95 | 2.20 | 19342 | $Co^{3+}$ | 3.529 | 14.47 |
| Fe LMM | 783.72 | 3.08 | 15241 | Fe Auger | | |
| Co 2p sat | 787.42 | 3.08 | 12465 | Satellite | | |
| Co 2p sat | 791.16 | 3.58 | 7264 | Satellite | | |
| Co 2p1/2 | 794.95 | 1.93 | 10640 | $Co^{2+}$ | | |
| Co2p1/2 | 795.92 | 1.68 | 9671 | $Co^{3+}$ | | |
| Co 2p sat | 797.38 | 3.08 | 6975 | Satellite | | |
| Co 2p sat | 803.47 | 3.08 | 6114 | Satellite | | |
| Ni 2p3 | 855.97 | 2.31 | 10212 | $Ni^{2+}$ | 2.309 | 7.74 |
| Ni 2p sat | 861.73 | 3.04 | 4714 | Satellite | | |
| Cu 2p3/2 | 933.14 | 2.94 | 25269 | $Cu^{2+}$ | 2.626 | 16.74 |
| Cu 2p sat | 940.87 | 2.06 | 2164 | Satellite | | |
| Mn LMM | 943.35 | 1.86 | 1836 | Mn Auger | | |

**Oxidation States Analysis:**

X-ray photoelectron spectroscopy (XPS) was performed in a ULVAC-PHI Inc. PHI GENESIS instrument equipped with monochromatic Al $K_\alpha$ radiation (1486.6 eV) as the excitation source. The X-ray analysis area for measurement was set at 100 μm diameter and a flood gun (1.6 V, 20 μA) was used for charge compensation. The pass energy was 112 eV for the high-resolution regions. The analysis chamber pressure was less than $1 \times 10^{-7}$ mbar during data acquisition. Fittings were performed with the MultiPak software. C1s C-C was used as a reference at 284.8eV for all spectra. Smart backgrounds were used to fit all elements. Constraints such as peak separation and area ratio confined to 2:1 for doublet peaks were used. Due to the high degree of transition metals with overlapping auger peaks, the peak fitting was nontrivial. For Mn, Ni and Cu, only the 2p3/2 peak area was fitted since they had no overlap with peaks outside the fitted region.

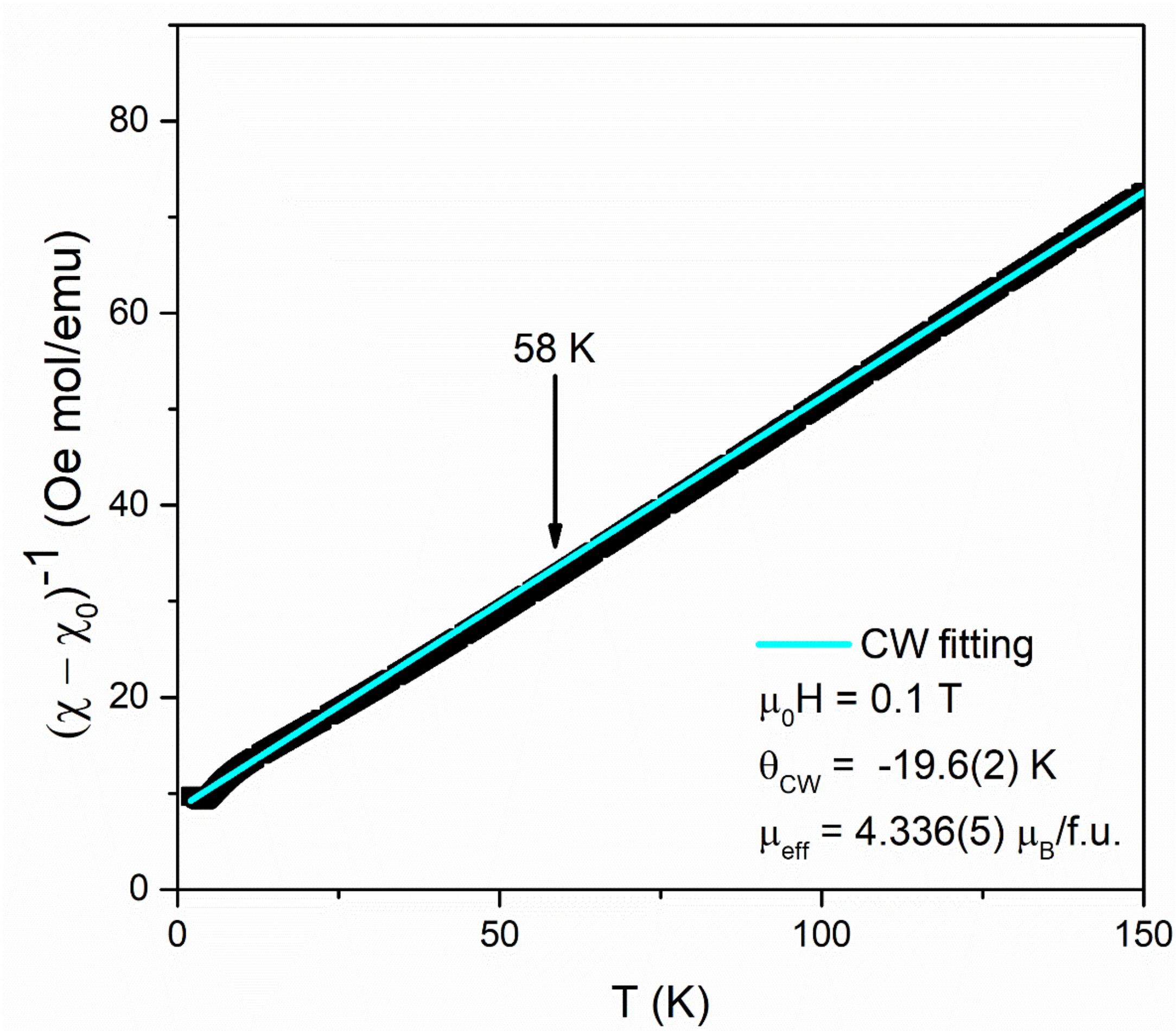

**Figure S4.** Deviation from the Curie-Weiss (CW) behavior at low temperatures.

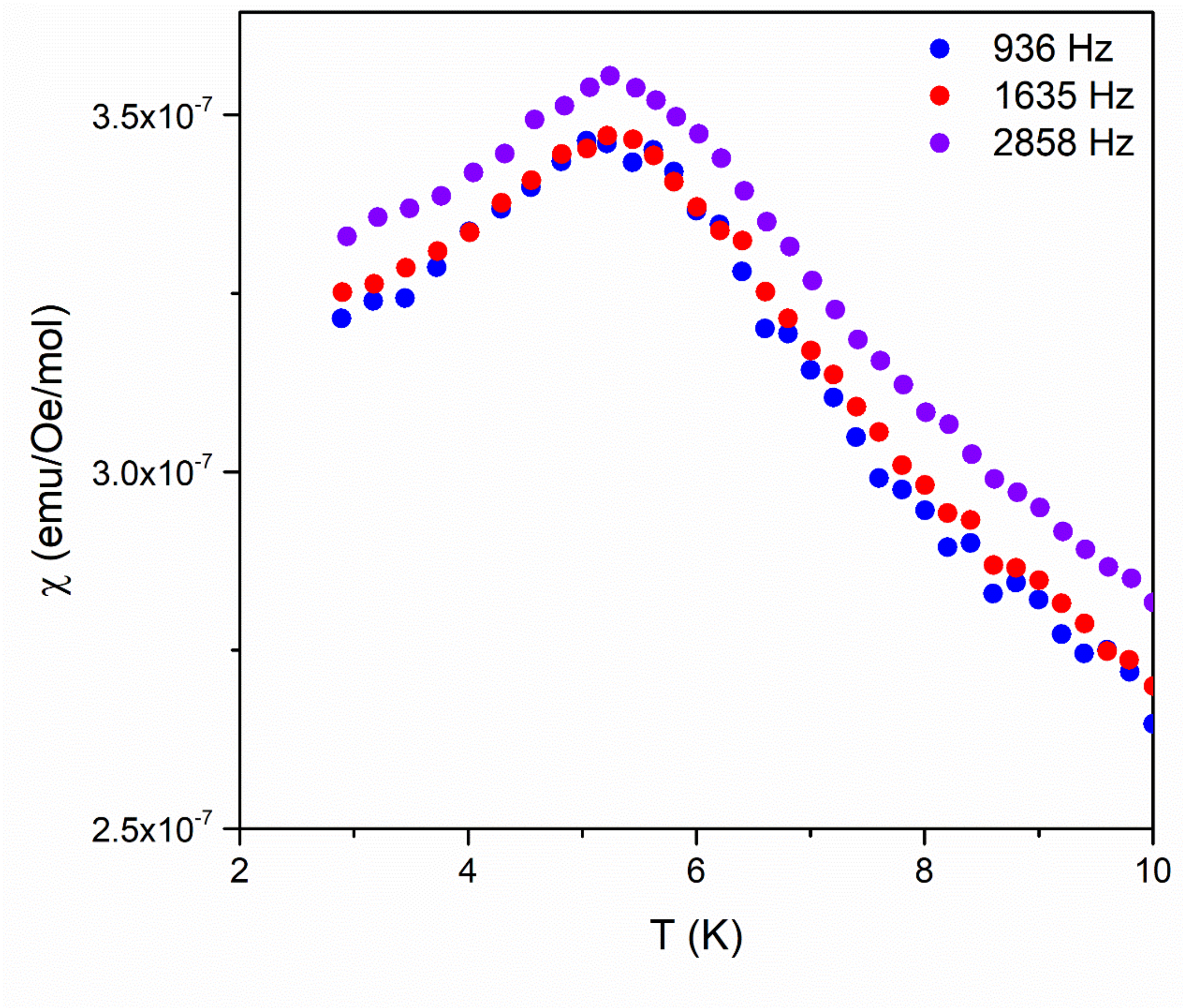

**Figure S5**. AC magnetic susceptibility of $(Mn_{0.2}Fe_{0.2}Co_{0.2}Ni_{0.2}Cu_{0.2})Ta_{1.92}O_{6-\delta}$ under various frequencies. The applied DC magnetic field is 100 Oe and the AC field is 10 Oe.

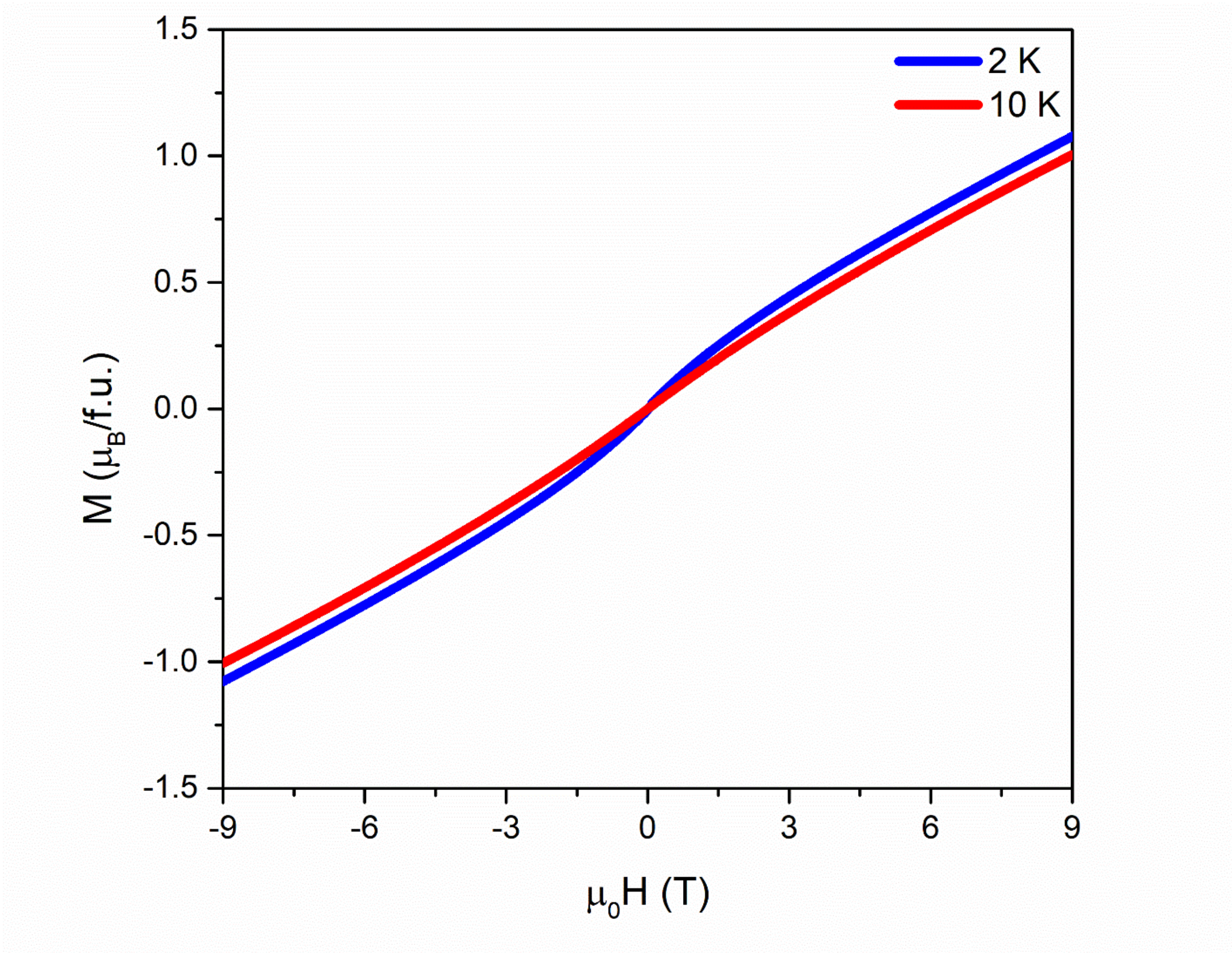

**Figure S6**. Hysteresis loops of $(Mn_{0.2}Fe_{0.2}Co_{0.2}Ni_{0.2}Cu_{0.2})Ta_{1.92}O_{6-\delta}$ at 2 K (blue) and 10 K (red).

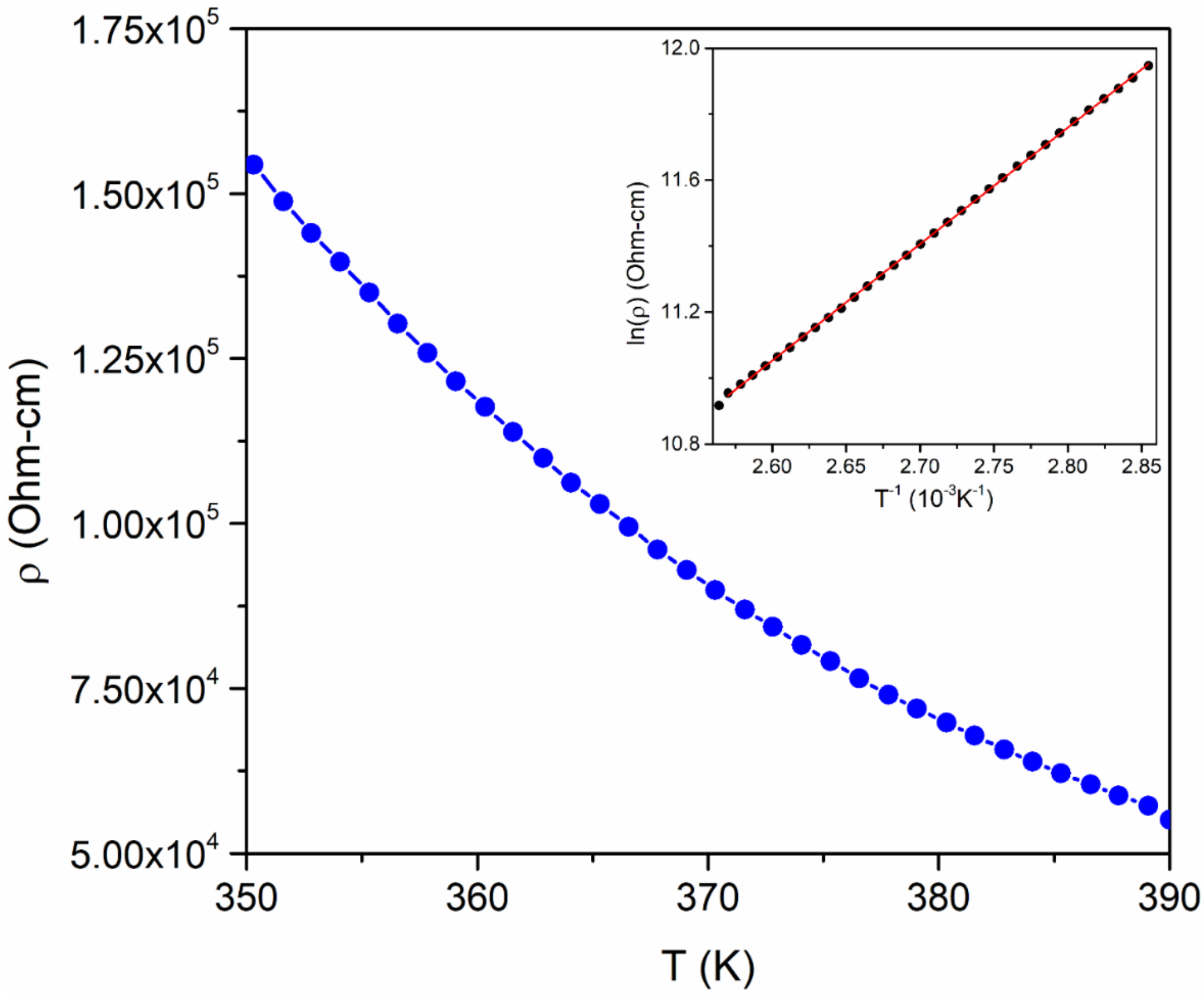

**Figure S7.** Resistivity of $(Mn_{0.2}Fe_{0.2}Co_{0.2}Ni_{0.2}Cu_{0.2})Ta_{1.92}O_{6-\delta}$ from 350 K to 390K displaying insulating behavior. The inset graph is of the linear relationship between ln(ρ) and $T^{-1}$. To calculate the electronic bandgap ($E_g$), the Arrhenius equation $\rho = \rho_0 e^{E_g/2k_BT}$, where $\rho_0$ is the pre-exponential term constant and $k_B$ is Boltzmann's constant, was used. $E_g$ was determined to be 608(1) meV.

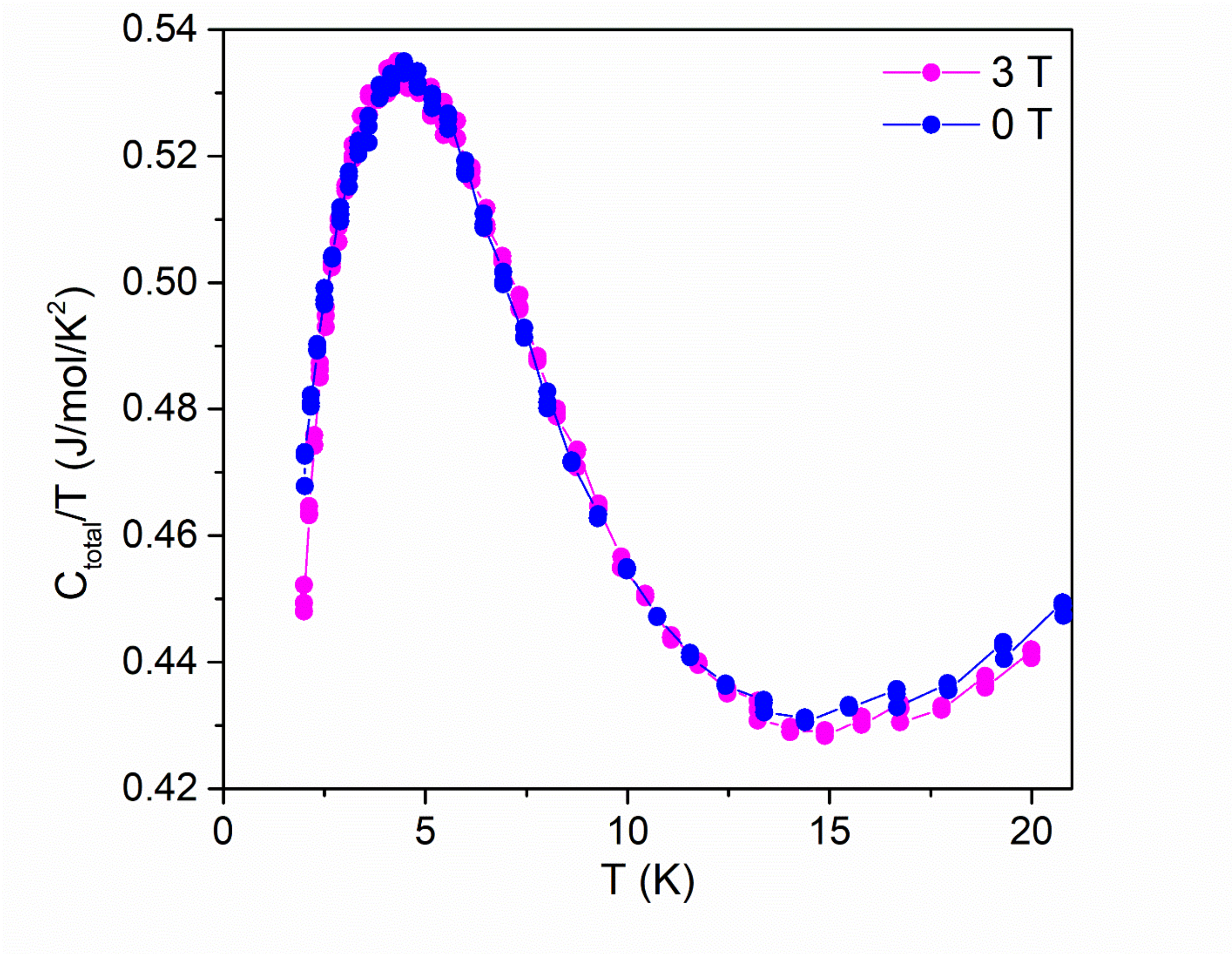

**Figure S8.** Temperature-dependent heat capacity of $(Mn_{0.2}Fe_{0.2}Co_{0.2}Ni_{0.2}Cu_{0.2})Ta_{1.92}O_{6-\delta}$ measured under magnetic field $\mu_0H$ = 0 T and 3 T.

**Molecular Orbital (MO) Calculation:** Semi-empirical extended-Hückel-tight-binding (EHTB) methods and CAESAR packages are used in calculating molecular orbitals of the parent compound, $CoTa_2O_6$.[1] The basis sets for Co are: 4*s*: Hii = -9.21 eV, ζ1 = 2, coefficient1 = 1.0000; 4*p*: Hii = -5.29 eV, ζ1 = 2, coefficient1 = 1.000; 3*d*: Hii = -13.18 eV, ζ1 = 5.55, coefficient1 = 0.568, ζ2 = 2.1, coefficient2 = 0.606. For Ta: 6*s*: Hii = -10.1 eV, ζ1 = 2.28, coefficient1 = 1.000; 6*p*: Hii = -6.86 eV, ζ1 = 2.241, coefficient1 = 1.000; 5*d*: Hii = -12.1 eV, ζ1 = 4.762, coefficient1 = 0.6815, ζ2 = 1.938, coefficient2 = 0.5589. For O: 2*s*: Hii = -32.29999 eV, ζ1 = 2.275, coefficient1 = 1.000; 2*p*: Hii = -14.8 eV, ζ1 = 2.275, coefficient1 = 1.000.